\documentclass[pra,twocolumn,10pt,superscriptaddress,showpacs]{revtex4-1}  %
\usepackage{graphicx,epsfig,epstopdf}
\usepackage[dvipsnames]{xcolor}
\usepackage{amsmath}
\usepackage{subfigure}
\usepackage{mathrsfs}
\usepackage{bm}
\usepackage{times}
\begin{document}

\title{Implementing universal nonadiabatic holonomic quantum gates with transmons}

\author{Zhuo-Ping Hong}
\affiliation{Guangdong Provincial Key Laboratory of Quantum Engineering
and Quantum Materials, and School of Physics\\ and Telecommunication Engineering,
South China Normal University, Guangzhou 510006, China}

\author{Bao-Jie Liu}
\affiliation{Department of Physics, South University of Science and Technology of China, Shenzhen 518055, China}
\affiliation{Guangdong Provincial Key Laboratory of Quantum Engineering
and Quantum Materials, and School of Physics\\ and Telecommunication Engineering,
South China Normal University, Guangzhou 510006, China}

\author{Jia-Qi Cai}
\affiliation{School of Physics, Huazhong University of Science and Technology, Wuhan 430074, China}


\author{Xin-Ding Zhang} \email{xdzhang@scnu.edu.cn}
\affiliation{Guangdong Provincial Key Laboratory of Quantum Engineering
and Quantum Materials, and School of Physics\\ and Telecommunication Engineering,
South China Normal University, Guangzhou 510006, China}

\author{Yong Hu} 
\affiliation{School of Physics, Huazhong University of Science and Technology, Wuhan 430074, China}

\author{Z. D. Wang}
\affiliation{Department of Physics and Center of Theoretical and Computational Physics, The University of Hong Kong, Pokfulam Road, Hong Kong, China}

\author{Zheng-Yuan Xue} \email{zyxue83@163.com}
\affiliation{Guangdong Provincial Key Laboratory of Quantum Engineering
and Quantum Materials, and School of Physics\\ and Telecommunication Engineering,
South China Normal University, Guangzhou 510006, China}

\date{\today}

\begin{abstract}
Geometric phases are well known to be  noise resilient in quantum evolutions and operations. Holonomic quantum gates provide us with a robust way towards universal quantum computation, as these quantum gates are actually induced by non-Abelian geometric phases. Here we  propose and elaborate how to efficiently implement universal nonadiabatic holonomic quantum gates on simpler superconducting circuits,  with a single transmon serving as a qubit. In our proposal, an arbitrary single-qubit holonomic gate can be realized in a single-loop scenario by varying the amplitudes and phase difference of two microwave fields resonantly coupled to a transmon, while nontrivial two-qubit holonomic gates may be generated with a transmission-line resonator being simultaneously coupled to the two target transmons in an effective resonant way. Moreover, our scenario may readily be scaled up to a two-dimensional lattice configuration, which is able to support large scalable quantum computation, paving the way for practically implementing universal  nonadiabatic holonomic quantum computation with superconducting circuits.
\end{abstract}


\maketitle

\section{Introduction}
As is known, a concept of phase factors is one of the most fundamental ones in quantum physics. In particular,  the state of a quantum system acquires a geometric phase in addition to the conventional dynamic one in a cyclic and adiabatic evolution~\cite{berry}.  As geometric phases are determined by the global properties of the evolution paths, they possess  a kind of built-in noise-resilience feature against certain types of local noises \cite{ps1,zhu05,ps2,mj}, which may naturally be used to achieve high fidelity quantum gates. For a practical larger system of qubits, the control lines and devices  inevitably induce local noises, and thus it is much more preferable to implement quantum gates in a geometric way. For this, considerable interest has been paid to various applications of geometric phases in quantum computation \cite{gqc}. Moreover,  due to the noncommutativity, non-Abelian geometric phases \cite{b3} can naturally lead to a universal set of quantum gates, i.e., the so-called holonomic quantum computation \cite{zanardi}.

On the side of physical implementation, schemes for quantum computation with non-Abelian geometric phases have been proposed for a variety of systems based on the adiabatic evolution with multilevel systems \cite{duan,adiabatic,adiabatic1,adiabatic2,adiabatic3, xdzhang,adiabatic4, adiabatic5,adiabatic6}, which appear to be rather complicated and thus difficult for experimental realization. Furthermore, the adiabatic condition requires the quantum dynamics to be slow, and thus decoherence effects may also introduce considerable errors \cite{xbwang,zhu}. Therefore, it is highly desirable to physically implement quantum gates with nonadiabatic evolutions \cite{b2}, in which the adiabatic condition is not required. For the past five years, significant research efforts have been devoted to the nonadiabatic holonomic quantum computation (NHQC) with three-level systems \cite{Sjoqvist2012, Xu2012, vam2014, Zhang2014d, Xu2014, Xu2015, Xue2015b, Xue2016, Zhao2016, Herterich2016,xu2017, vam2017, Xue2017,xu20172,zhao2017,vam2016}, according to which fast holonomic quantum gates may be obtained with simpler physical systems. Notably, such a nonadiabatic idea has experimentally been demonstrated in superconducting circuits \cite{Abdumalikov2013}, NMR \cite{Feng2013,li2017}, and  electron spins in diamond \cite{Zu2014, Arroyo-Camejo2014, nv2017, nv20172}.

Due to the good flexibility and scalability, superconducting quantum circuits \cite{sq1,sq2,sq3,sq4} have been one of the promising platforms for implementing quantum computation. Recently, high energy levels of a superconducting  transmon qubit \cite{transmon} were also shown to possess long coherence times \cite{multilevel}, which means transmons can also be used as multilevel quantum systems.  However, the spectrum of transmons is weakly anharmonic and thus leads to spectral crowding in multiqubit scenarios where qubit-qubit interactions are induced by dispersive couplings between transmission-line resonators (TLRs) and transmon qubits \cite{Xue2015b}.  Also, as to NHQC, the complicated interaction needed for a nontrivial two-qubit holonomic gate \cite{Sjoqvist2012} is still experimentally challenging, and thus only single-qubit gates have been achieved experimentally \cite{Abdumalikov2013}. Therefore,  it is extremely desirable to implement the two-qubit gates using only the simple resonant TLR-qubit interaction.

Here we propose and elaborate on how to implement {\it universal} NHQC using superconducting circuits, which removes the above-mentioned difficulties. In the current implementation, each transmon serves as a qubit, as in the experiment of Ref. \cite{Abdumalikov2013}. An arbitrary holonomic single-qubit gate can be obtained by varying the amplitudes and phase difference of two microwave fields resonantly coupled to a transmon, where an arbitrary single-qubit gate can be realized by using a single implemented gate, i.e.,  a single cyclic evolution. Therefore, it is an extension of the nonadiabatic holonomic single-qubit gates demonstrated in  Ref. \cite{Abdumalikov2013}, which combines two sequential gates to obtain an arbitrary single-qubit gate, and thus is essentially different from our implementation. Moreover,  nontrivial two-qubit gates can be achieved with a TLR being simultaneously coupled to the two qubits driven by microwave fields, in which the effective resonant tunable TLR-qubit couplings are induced, as in Ref. \cite{jc}, for the two involved qubits. In this configuration, the three coupled quantum system can also be seen as a three-level system in the single-excitation subspace, and thus can be directly used to induce nontrivial  two-qubit gates, in analogy to the single-qubit case.
In addition, the present scenario can  readily be scaled up to a two-dimensional lattice for scalable quantum computation.


\begin{figure}[tbp] \centering
\includegraphics[width=8cm]{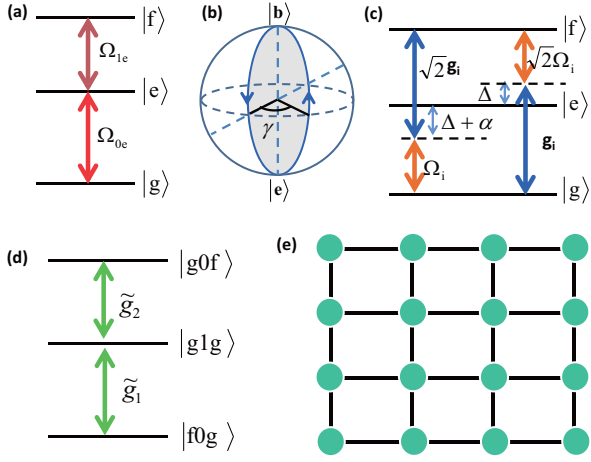}
\caption{Illustration of the proposed scheme. (a) The coupling configuration for the single-qubit gates with two microwave fields resonantly coupled to the three levels of a transmon qubit. (b) Geometric illustration of the proposed single-qubit gate. (c)  For nontrivial two-qubit gates, both qubits are coupled a TLR and driven by microwave fields, which induced effective resonant qubit-TLR coupling. (d) In this single-excitation subspace, the effective coupling configuration for the two qubits and the TLR for nontrivial two-qubit gates. (e) Scale-up of our scheme, where the transmon qubits and TLRs are denoted by filled circles and bonds, respectively.}\label{setup}
\end{figure}

\section{Universal single-qubit gates}
We now proceed to present our scheme. We first address how to implement an arbitrary single-qubit gate, as shown in Fig.~\ref{setup}(a), which consists of two microwave fields with amplitudes $\Omega_{ie}(t)$ ($i=0, 1$) and initial phases $\phi_i$ resonantly coupled to the sequential transitions of the three lowest levels $|g\rangle$, $|e\rangle$, and $|f\rangle$ of a transmon, with $|g\rangle$ and $|f\rangle$ being the qubit states. The Hamiltonian of the system may be written as
\begin{eqnarray} \label{h1}
H_{1} &=& \Omega_{0e}(t)e^{i\phi_0}|g\rangle\langle e|+
e^{i\phi_1}\Omega_{1e}(t)|f\rangle\langle e|  + \mathrm{H.c.}\\
&=&\Omega (t) e^{i(\phi_1-\pi)} \left(\sin\frac{\theta}{2}e^{i\phi}|g\rangle -\cos\frac{\theta}{2}|f\rangle\right)\langle e| + \mathrm{H.c.},\notag
\end{eqnarray}
where $\phi=\phi_0-\phi_1+\pi$, $\Omega(t)=\sqrt{\Omega_{0e}(t)^2+\Omega_{1e}(t)^2}$, and $\tan(\theta/2)=\Omega_{0e}(t)/\Omega_{1e}(t)$. That is, the dynamics of the system is captured by the resonant coupling between the  states $|b\rangle=\sin(\theta/2)e^{i\phi}|g\rangle-\cos(\theta/2)|f\rangle$ and $|e\rangle$, while the dark eigenstate $|d\rangle=\cos(\theta/2)|g\rangle+\sin(\theta/2)e^{i\phi}|f\rangle$ is left unchanged. Therefore, when the cyclic evolution condition is met, i.e., $\int_0^T \Omega(t) dt=\pi$, one can obtain a certain  single-qubit gate  by choosing different $\theta$ and/or $\phi$. Meanwhile,  as $\langle j(t)|H_1|i(t)\rangle=0$ with $i\in\{b, d\}$,  there are no transitions between the $|d\rangle$ and $|b\rangle$ states during the evolution (meet the parallel-transport condition), and the dynamical phases for the $|d\rangle$ and $|b\rangle$ states are also zero. Therefore, the obtained single-qubit gates are of the holonomic nature \cite{Sjoqvist2012}.

To achieve the set of universal single-qubit gates in a more general scenario, we set the total evolution time $T$ to be divided into two equal intervals, i.e., $\phi_0=0$, $\phi_1= \pi$ for $t\in [0, T/2]$, and $\phi_0'=\pi+ \gamma$,  $\phi_1'= \gamma$  for $t\in [T/2, T]$. In this case, the Hamiltonians that dominate the two consecutive evolution paths are $H_a=\lambda_1(|b\rangle\langle e|+ |e\rangle\langle b|)$ and $H_b=-\lambda_1 (e^{i\gamma} |b\rangle\langle e|+ e^{-i\gamma} |e\rangle\langle b|)$, and the corresponding evolution operators are, respectively, $U_a=|d\rangle\langle d| -i (|b\rangle\langle e|+ |e\rangle\langle b|)$ and $U_b=|d\rangle\langle d| + i (e^{i\gamma} |b\rangle\langle e|+ e^{-i\gamma} |e\rangle\langle b|)$. As a result, the single-qubit gate operator is given by
\begin{eqnarray}\label{singleloop}
U_1(\theta,\phi) &\sim& \left(\begin{array}{cc}
\cos \frac{\gamma}{2}-i\sin\frac{\gamma}{2}\cos\theta & -i\sin\frac{\gamma}{2}\sin\theta e^{i\phi}\\
-i \sin\frac{\gamma}{2}\sin\theta e^{-i\phi} &\cos\frac{\gamma}{2}+i\sin\frac{\gamma}{2}\cos\theta
\end{array}\right) \nonumber \\
&=& \exp\left({-i{\gamma \over 2} {\bf n}\cdot{\bf \sigma}}\right),
\end{eqnarray}
which describes a rotation operation around the axis ${\bf n}= (\sin\theta\cos\phi, \sin\theta\sin\phi, \cos\theta)$ by an angle $\gamma/2$ and can generate the set of universal single-qubit gates in the qubit subspace, up to a global phase factor $\exp(i\gamma/ 2)$, in a holonomic way \cite{Herterich2016}. For instance, for the two different set of $(\theta, \phi)$, the corresponding two sets of gates (with different $\gamma$) form a universal set of single-qubit ones. Also  from a geometrical point of view, the above two Hamiltonians correspond to two different  paths in the Bloch sphere in a consecutive and cyclic way: the final point (at $T/2$) of $H_a$ is coincident with the start point (at $T/2$) of $H_b$, while the final point (at $T$) of $H_b$ is just the starting point of $H_a$ (at 0). That is, the two paths coincide at 0 and $T/2$, with the cyclic geometric phase being illustrated as the slice contour in Fig. \ref{setup}(b). We want to emphasize that $\Omega(t)$ in our single-loop scheme can be in an arbitrary shape, providing that the two microwave fields are in the same shape. This is due to the fact that our scheme is a resonant one, which is a merit compared with the two most recent experiments \cite{nv2017,nv20172} with detuning. In the detuned schemes, the detuning to the auxiliary state should also be in the same shape as the driven fields, which makes the experiment more difficult as the frequencies of driven fields must also be changed in order to change the detuning. Therefore, the  experiments \cite{nv2017,nv20172} are done with square pulses to avoid tuning the detuning. In this case, an ideal square pulse is needed, which is difficult for a large-amplitude pulse as the pulse will be very sharp, and thus leads to infidelity \cite{nv20172}.

\begin{figure}[tbp]\centering
\includegraphics[width=8cm]{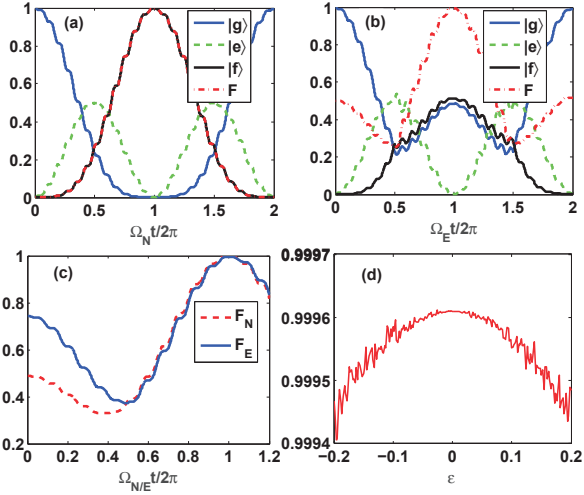}
\caption{State population and fidelity dynamics of different $(\gamma, \theta)$ gates,  (a) $(\pi, \pi/2)$ and (b) $(\pi/2, \pi/2)$, as a function of $\Omega_{N/E}t/\pi$ with the initial state being  $|g\rangle$. (c) Dynamics of the gate fidelities. (d) The stability of the NOT gate to certain random fluctuations $\epsilon\Omega$. }
\label{Fig f1}
\end{figure}

The performance of a single-qubit gate in Eq. (\ref{singleloop}) can be evaluated by using the following quantum master equation:
\begin{eqnarray}  \label{me}
\dot\rho_1 &=& i[\rho_1, H_1]  + \frac{1}{2} \sum_{j\in\{g, f\}} \left[ \Gamma_1^{j} \mathcal{L}(\sigma^-_{j,e}) + \Gamma_2^{j} \mathcal{L}(\sigma^\text{z}_{j,e})\right],
\end{eqnarray}
where $\rho_1$  is the density matrix of the considered system and $\mathcal{L}(A)=2A\rho_1 A^\dagger-A^\dagger A \rho_1 -\rho_1 A^\dagger A$ is the Lindbladian of the operator $A$, $\sigma^-_{g,e}=|g\rangle\langle e|$, $\sigma^-_{f,e}=|e\rangle\langle f|$, $\sigma^\text{z}_{f,e}= (|f\rangle\langle f|-|e\rangle\langle e|)$, and $\sigma^\text{z}_{g,e}= (|e\rangle\langle e|-|g\rangle\langle g|)$. In addition, $\kappa$, $\Gamma_1^j$,  and $\Gamma_2^j$ are the decay rate of the cavity, the decay and dephasing rates of the $\{j,e\}$ two-level systems, respectively. Suppose that the qubit is initially in the state $|g\rangle$. We then evaluate the NOT gate of $\gamma=\pi,\theta=\pi/2$  and the gate with $\gamma=\pi/2, \theta=\pi/2$, using the fidelity defined by $F=\langle\psi_f|\rho_1|\psi_f\rangle$ with  $|\psi_f\rangle=|f\rangle$ and $|\psi_f\rangle=[(1+i)|g\rangle+(1-i)|f\rangle]/2$ being the corresponding target state. The obtained fidelities are as high as $F _{N}= 99.75\%$ and $F _{E}= 99.56\%$ at $t=\pi/\Omega_{N/E}$, as shown in Figs. \ref{Fig f1}(a) and \ref{Fig f1}(b). The infidelity is mainly due to relaxation and dephasing of the qubits and the resonator. The parameters of the qubit are set as $\Omega(t) = 2\pi\times 16$ MHz,  and $\Gamma_1^j=\Gamma_2^j= \kappa=2\pi \times 10$ kHz,  corresponding to the coherent time of 16 $\mu$s, which is easily accessible with current technologies \cite{multilevel}. The anharmonicity of the third level is set to be $\alpha=\omega_{ge}-\omega_{fe}=2\pi\times 400$ MHz \cite{sq4}. We modulate $\Omega_{0e}=\Omega_{1e}= \Omega_{N/E}/\sqrt{2}$ to ensure $\theta = \pi/2$. In addition, for a general initial state of $|\psi\rangle=\cos\theta^{'}|g\rangle+\sin\theta^{'}|f\rangle$ where $\theta^{'}=0$ corresponds to the $|0\rangle$ state, we have numerically confirmed that the fidelity changes slightly when $\theta^{'}>0$. Therefore, to fully quantify the performance of the implemented gate, in Fig. \ref{Fig f1}(c) we have plotted the gate fidelities for 1001 input states with $\theta^{'}$ uniformly distributed over $[0,2\pi]$, where we find that $F _{N}= 99.82\%$ and $F _{E}= 99.57\%$. On the other hand,  we numerically  demonstrate that the gate is also insensitive to certain random fluctuations with relatively high frequencies. The randomized fluctuation is artificially introduced by adding an amplitude shift to $\Omega$ as $\Omega^{'}=(1+\epsilon)\Omega$, where $\epsilon$ has 1000 points of noise and a mean value of zero. In the absence of decoherence, the fidelity of the NOT gate  is almost stable at 1 when $\epsilon$ increases up to $20\%$.

\section{Nontrivial two-qubit gates}
At this stage, we turn to the implementation of nontrivial two-qubit gates, where we consider the case that the two driving qubits are coupled simultaneously to a nonlinear TLR. As illustrated in Fig. \ref{setup}(c), both the TLR and the driving microwave field are dispersively coupled with the transitions $|g\rangle\leftrightarrow|e\rangle$ and $|f\rangle\leftrightarrow|e\rangle$ with frequencies  $\omega_{ge}$ and  $\omega_{fe}$, the coupling strength for the $i$th qubit being $g_i$ and $\Omega_i$ and their corresponding frequencies being $\omega_c$ and $\omega_i$. Meanwhile, the two couplings form a two-photon resonant situation, i.e., $\omega_c-\omega_{ge}=\omega_{fe}-\omega_i=\Delta>\alpha$, and thus lead to driving-assisted coherent resonant coupling between the TLR and the $|g\rangle\leftrightarrow|f\rangle$ transition \cite{jc}. When $\Delta \gg \{g_i, \Omega_i\}$, after concealing the ac Stark shifts by modulating the frequencies of the driven fields accordingly to $\Omega_i$ (see Fig. 4 in Appendix B), the interacting system can be written as
\begin{eqnarray}\label{coupled}
H_{2} &=& \sum_{i=1}^2 \tilde{g}_i(e^{-i\varphi_i} a |f\rangle_i\langle g| + \mathrm{H.c.}),
\end{eqnarray}
where $\tilde{g}_i =\sqrt{2}g_i\Omega_i\alpha/[\Delta(\Delta-\alpha)]$ and $\varphi_i$ is the initial phase of the microwave driving field on $i$th qubit, see Appendix A for details. Note that, as $\Delta$ and $\alpha$ are comparable, the effective interaction is obtained from the interference of the two paths as illustrated in \ref{setup}(c). In addition, we note that one can obtain stronger $\tilde{g}_i$  by enlarging  its corresponding $\Omega_i$. However,  when $\Omega_i$ is large, the linear dependence of  $\tilde{g}_i$ with respective to $\Omega_i$ will no longer hold, as the perturbation theory is not good then. However, we can still get the $\tilde{g}_i$ and $\Omega_i$ correspondence numerically, as shown in Fig. 5 in Appendix B.

We now show that the Hamiltonian of Eq. (\ref{coupled}) can readily be employed to implement nontrivial two-qubit gates. In the single-excitation subspace $S_1=\mathrm{span}\{|f0g\rangle, |g0f\rangle, |g1g\rangle\}$, where $|jnk\rangle \equiv |j\rangle_1\otimes|n\rangle_c\otimes|k\rangle_2$ labels the product states of the two qubits and the TLR, Eq. (\ref{coupled}) may be rewritten as
\begin{eqnarray}  \label{h2}
H_\mathrm{eff}=g  \left(\sin\frac{\vartheta}{2} |f0g\rangle
-\cos\frac{\vartheta}{2}|g0f\rangle\right)\langle g1g| + \mathrm{H.c.},
\end{eqnarray}
where $g=\sqrt{\tilde{g}_1^2+\tilde{g}_2^2}$, $\tan(\vartheta/2)
= \tilde{g}_1/\tilde{g}_2$. In the derivation, we have also set $\varphi_1=0$ and $\varphi_2=\pi$, i.e., a $\pi$ difference for the initial phases of the two driving microwave fields. Equation (\ref{h2}) establishes a coupled three-level Hamiltonian in the single-excitation subspace, with the TLR excitation state $|g1g\rangle$ to serve as an ancillary state, as shown in Fig. \ref{setup}(d), being the same as that of single-qubit gate case. Therefore, holonomic quantum gates can be obtained for the two-qubit states $|f0g\rangle$ and $|g0f\rangle$, which are the odd parity subspace $\{|gf\rangle, |gf\rangle\}$ when neglecting the states of the TLR (always to be the vacuum state after a gate operation). We want to emphasize that our construction of the two-qubit gate involves only a single three-level structure in the two-qubit Hilbert space, which is simper than that in Ref. \cite{Sjoqvist2012}, where the two-qubit gates need two three-level structures. The dynamics under Hamiltonian (\ref{h2}) can be captured by a resonant coupling between the bright state $|b\rangle_2=\sin(\vartheta/2)|f0g\rangle-\cos(\vartheta/2)|g0f\rangle$ of Hamiltonian (\ref{h2}) and the ancillary state $|g1g\rangle$, with the effective Rabi frequency $g$, while the dark state $|d\rangle_2=\cos(\vartheta/2)|f0g\rangle +\sin(\vartheta/2)|g0f\rangle$ is decoupled. When $\int_0^\tau g dt=\pi$, the dressed states undergo a cyclic evolution, with $|b \rangle_2$ evolving to $-| b \rangle_2$ and $| d \rangle_2$ remaining unchanged. Moreover, as $\langle\psi_{\text{i}}(t)|H_1|\psi_{\text{j}}(t)\rangle=0$ with $|\psi_{\text{i,j}}\rangle\in \{|d\rangle_2, |b\rangle_2 \}$, the evolution satisfies the parallel-transport condition and acquires no dynamical phases. Thus, the evolution operator $U_2=\exp(-i\int_0^{\tau} H_2 dt)$ realizes holonomic operations.  In the two-qubit gate Hilbert space $S_2=\text{span}\{|gg\rangle,|fg\rangle,|gf\rangle, |ff\rangle\}$, the corresponding gates are
\begin{equation}\label{u2}
U_2(\vartheta)= \left(\begin{array}{cccc}
1&0&0&0\\
0&\cos{\vartheta}&\sin{\vartheta}&0\\
0&\sin{\vartheta}&-\cos{\vartheta}&0\\
0&0&0&-1\\
\end{array}\right),
\end{equation}
which induces only a kind of nontrivial transformation to the odd parity subspace of $S_2$, as expected, and thus implements nontrivial two-qubit gates. Meanwhile, the minus sign of the $|ff\rangle\langle ff|$ elements comes from the evolution of the dual two-excitation subspaces of $\{|g1f\rangle, |f0f\rangle, |f1g\rangle \}$.

\begin{figure}[tbp]\centering
\includegraphics[width=6.5cm]{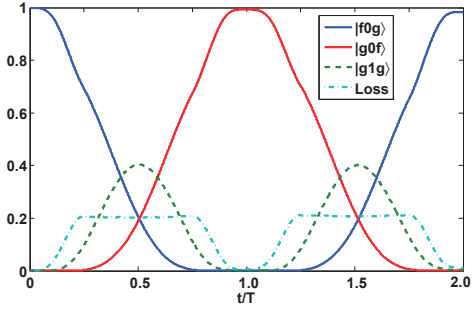}
\caption{State population and fidelity dynamics for   gates as a function of time with the initial state being  $|f0g\rangle$. }
\label{f2}
\end{figure}

We now analyze the performance of two-qubit gates. For $\Delta=2\pi\times 1$ GHz, $g_i=2\pi \times 65$ MHz \cite{jc}, one can obtain
\begin{eqnarray} \label{pulse}
\tilde{g}_i=g_0\times
  \begin{cases}
    \sin^2\left(2\pi t/T\right), &  0 \leq t < T/4; \\
    1, &  T/4 \leq t \leq 3T/4; \\
    \sin^2\left[2\pi (T-t)/T\right], & 3T/4 < t \leq T;
  \end{cases}
\end{eqnarray}
with $T=40$ ns, $g_0\simeq 2\pi \times 11.8$ MHz by modulating $\Omega_i$ with the maximum value to be $2\pi\times 377$ MHz. In the case of $\vartheta=\pi/2$, the induced two-qubit gate is the SWAP-like gate for the two qubits. When the initial state of the two-qubit state is $|fg\rangle$, as shown in Fig. \ref{f2}, a fidelity $F_2=99.44\%$ can be obtained. We want to emphasize that, the simulations must be done faithfully based on the original Hamiltonian, i.e., including the unwanted higher-order effects induced by the strong microwave drive, such as off-resonant transitions to higher transmon levels. We also note that there is loss from our computational basis, which is due to the time dependence of the amplitude of the pulse in Eq. (\ref{pulse}), leading to the time dependence of the  ac Stark shift terms, which can also be compensated by modulating  the pulse frequencies $\omega_i$ accordingly.  However, as we conceal the energy shifts, the loss is zero before and after the operation and thus leads to high fidelity gates.

\section{Discussion and conclusion}
Our proposal allows physical realization of universal nonadiabatic holonomic quantum gates, i.e., single-qubit gates on transmon qubits from individual control, and two-qubit gates induced between any two qubits sharing the same TLR serving as ancillary. As to the experimental feasibility of our proposal, it is noted that the elementary gates require the  transmon qubits and the TLRs to be individually controlled. Considering that both the dc and ac flux controls in coupled superconducting qubits have already been achieved \cite{flux1,flux2}, where the qubit loop sizes and their distances are on the order of micrometers, the individual control of our scheme is quite feasible with the current technologies. Moreover, the present scheme may readily be scaled up to a two-dimensional lattice configuration by placing the TLRs and transmon qubits in an interlaced square lattice, as shown in Fig. \ref{setup}(e), and thus facilitating the scalability criteria of quantum computation. Experimentally, a small  lattice of this type has already been demonstrated \cite{lattice}. As for a large-scale lattice, the individual control, wiring, and readout can be conveniently integrated in an additional layer on top of the  qubit lattice layer \cite{barends,yale1,yale2}, and the interlayer connection may be achieved by capacitive coupling. Finally, we wish to note that our scheme is insensitive to the background charge noise as it is made of only TLRs and the charge-insensitive transmons \cite{transmon}. For the flux type and critical current type $1/f$ noise,  the influence is even weaker than the intrinsic decay effect \cite{noise}, which has already been considered within our numerical simulations.

To conclude, we have proposed and elaborated how to efficiently implement universal NHQC with superconducting transmon qubits with resonant coupling. Meanwhile, our proposal can be scalable. It is anticipated that the present simpler and more efficient scheme will stimulate significant experimental interest for realizing it, paving the way for implementing robust NHQC using superconducting circuits.

\acknowledgements
This work was supported in part by the National Key R\&D Program of China (Grant
No. 2016YFA0301802 and No. 2016YFA0301803) and National Natural Science Foundation  of China (Grant
No. 11774114).

Z.P.H., B.J.L. and J.Q.C. contributed equally to this work.

\appendix

\section{The effective Hamiltonian}
We consider the effective Hamiltonian of Eq. (4) in the main text, which deals with the effective Hamiltonian in a projective subspace of the original Hamiltonian. For the following Hamiltonian of the $i$th transmon coupled to the TLR,
\begin{equation}
	\begin{gathered}
	{H_0} = {\delta _r}{n_a} + {\delta _q}{n_b} - \frac{\alpha }{2}\left( {{n_b} - 1} \right){n_b} \hfill \\
	H' = ga{b^\dag } + \frac{{\Omega {e^{i\phi }}}}{2}b + \text{H.c.}, \hfill \\
	\end{gathered}
	\end{equation}
where $\delta_r=\omega_c-\omega$ with $\omega$ being the frequency of the drive microwave field, $\delta_q=\omega_{g,e}-\omega$, and $n_a = a^{\dagger} a, n_b = b^{\dagger}b$. With $b=|g\rangle \langle e|+\sqrt{2}|e \rangle \langle f|+\sqrt{3}|f \rangle \langle h|+...$ being the lower operator for the transmon, the energies of the state $\left| {g,1} \right\rangle ,\left| {f,0} \right\rangle $ are
\begin{equation}
{E_{f,0}} = 2{\delta _q} - \alpha,  \quad
{E_{g,1}} = {\delta _r},
\end{equation}
which can be adjusted to be degenerate by modulating $\omega_i$ such that ${\delta _r} = 2{\delta _q}- \alpha$.

We then define
\begin{eqnarray}
    P &=& \left| {g,1} \right\rangle \left\langle {g,1} \right| + \left| {f,0} \right\rangle \left\langle {f,0} \right|,   \\
    K&=& \sum\limits_\pi  {\frac{{\left| {i,n} \right\rangle \left\langle {i,n} \right|}}{{{\varepsilon _{i,n}} - \varepsilon }}},
\end{eqnarray}
where the subspace $P$ is of interest and $\pi :\left\{ {\left. {i,n} \right|\left( {i,n} \right) \ne \left( {g,1} \right)or\left( {f,0} \right)} \right\}$. In the following calculation, we restrict ourselves within the qubit subspace of the first four levels, as we only involve the first three levels in our gate implementation.

Once we handle the effective Hamiltonian using a perturbation theory with $\{g, \Omega_i\} \ll \Delta=\delta_q-\delta_r$, the first-order term is found to be
\begin{equation}
    {{\tilde H}_1} = PH'P=0,
\end{equation}
as
\begin{eqnarray}
   && H'P = \left( { g\left| {e,0} \right\rangle  + \frac{\Omega }{2}{e^{ - i\phi }}\left| {e,1} \right\rangle } \right)\left\langle {g,1} \right|  \\
    & +&\left( {\sqrt 2 g\left| {e,1} \right\rangle  + \frac{{\sqrt 2 \Omega }}{2}{e^{i\phi }}\left| {e,0} \right\rangle  + \frac{{\sqrt 3 {e^{ - i\phi }}}}{2}\left| {h,0} \right\rangle } \right)\left\langle {f,0} \right|. \notag
\end{eqnarray}

As for the second-order terms,
\begin{eqnarray}
{{\tilde H}_2} &=& - PH'KH'P \notag \\
    &=& - PH'\left( K_1 + K_2 + K_3 \right)H'P,
    \end{eqnarray}
 where
    \begin{eqnarray}
K_1 &=& \frac{{\left| {e,0} \right\rangle \left\langle {e,0} \right|}}{{{\varepsilon _{e,0}} - \varepsilon }},
K_2 = \frac{{\left| {e,1} \right\rangle \left\langle {e,1} \right|}}{{{\varepsilon _{e,1}} - \varepsilon }},
K_3 = \frac{{\left| {h,0} \right\rangle \left\langle {h,0} \right|}}{{{\varepsilon _{h,0}} - \varepsilon }}. \notag
    \end{eqnarray}
Finally, we have
\begin{eqnarray}\label{jc}
{{\tilde H}_2} &=& {\eta_{g1}}\left| {g,1} \right\rangle \left\langle {g,1} \right|
+ {\eta_{f0}}\left| {f,0} \right\rangle \left\langle {f,0} \right|\notag\\
&&+ \left( \tilde{g} \left| {f,0} \right\rangle \left\langle {g,1} \right| + \text{H.c.} \right),
\end{eqnarray}
where
\begin{eqnarray}
{\eta_{g1}}&=& \frac{{{g^2}}}{\Delta } -\frac{{{\Omega ^2}}}{{4\left( {\Delta+\alpha }\right)}}\notag \\
{\eta_{f0}} &=&\frac{{{\Omega ^2}}}{{2\Delta }}-\frac{{2{g^2}}}{{\Delta  + \alpha }}
                    -  \frac{{3{\Omega ^2}}}{4({{\Delta} - \alpha })}\\
\tilde{g}  &=& \frac{{\sqrt 2g\Omega {e^{ - i\phi }}}}{2\Delta } -\frac{{ \sqrt 2 g\Omega {e^{ - i\phi }}}}{2(\Delta  + \alpha)}
    = \frac{{g\Omega {e^{-i\phi }}\alpha }}{{\sqrt 2 \Delta \left( {\Delta  + \alpha } \right)}}.\notag
\end{eqnarray}

\begin{figure}[tbp] \centering
\includegraphics[width=7cm]{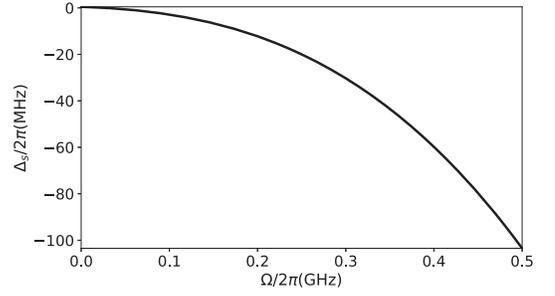}
\caption{Illustration of the ac Stark shift to be compensated for fixed $g$ with respect to $\Omega$.}\label{ac}
\end{figure}

\section{Compensate of the ac stack}

In Eq. (\ref{jc}), there are ac Stark shifts, which will lead to $\sim$ 3\% infidelity of the gate operations. Therefore, we need to compensate these shifts. It is noted that both $g$ and $\Omega$ split the degenerate subspace, so we will fix $g$ and tune the frequency $\omega$ of the driven field to let $|\eta_{g1g1}-\eta_{f0f0}|=0$, i.e.,
$$\left\langle {{\phi _j}\left( \Omega  \right)} \right|\frac{d}{{d\Omega \left( t \right)}}\left| {{\phi _i}\left( \Omega  \right)} \right\rangle  = 0.$$
As $$\left\langle {{\phi _j}}(\Omega ) \right|\frac{d}{{d\Omega }}\left| {{\phi _i}}(\Omega ) \right\rangle
=  {\left\langle {{\phi _j}} (\Omega )\right|\frac{{dH}}{{d\Omega }}\left| {{\phi _i}} (\Omega ) \right\rangle }/({{E_i} - {E_j}}),$$
we obtain
$$\left\langle {{\phi _j}} (\Omega )\right|\frac{{\partial H}}{{\partial \Omega }}\left| {{\phi _i}} (\Omega )\right\rangle  + \frac{{d\omega }}{{d\Omega }}\left\langle {{\phi _j}} (\Omega ) \right|\frac{{\partial H}}{{\partial \omega }}\left| {{\phi _i}} (\Omega ) \right\rangle  = 0,$$
which can be numerically solved to obtain the $\omega-\Omega$ curve, such that one can figure out the ac Stark shift $\Delta_s$ to be compensated, as shown in Fig 4.

\begin{figure}[tbp] \centering
\includegraphics[width=6.8cm]{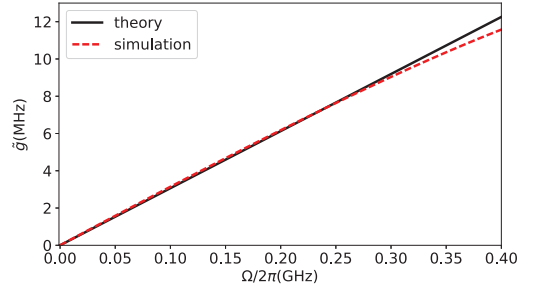}
\caption{Illustration of the effective transmon-TLR coupling strength with respect to $\Omega$, with fixed $g$.}\label{geff}
\end{figure}
Therefore, to effectively conceal the ac Stark shifts, we need a driven pulse with smoothly changed amplitude, which leads to  a smoothly changed effectively resonant coupling strength $\tilde{g}$. Also note that when $\Omega$ is large, the $\tilde{g_i}-\Omega$ dependence will be slightly nonlinear, as shown by the red dashed line in Fig. 5, while the result of perturbation theory is indicated by the black solid line.


\begin{thebibliography}{99}

\bibitem{berry} M. V. Berry, Proc. R. Soc. Lond., Ser. A {\bf 392}, 45 (1984).

\bibitem{ps1} P. Solinas, P. Zanardi, and N. Zangh\`{\i},
Phys. Rev. A {\bf 70}, 042316 (2004).

\bibitem{zhu05} S.-L. Zhu and P. Zanardi,
   Phys. Rev. A {\bf 72}, 020301(R) (2005).

\bibitem{ps2} P. Solinas, M. Sassetti, T. Truini, and N. Zangh\`{\i},
New J. Phys. {\bf 14}, 093006 (2012).

\bibitem{mj} M. Johansson, E. Sjoqvist, L. M. Andersson, M. Ericsson,
B. Hessmo, K. Singh, and D. M. Tong,
Phys. Rev. A {\bf 86}, 062322 (2012).

\bibitem{gqc} E. Sj\"{o}qvist, Physics {\bf1}, 35 (2008).

\bibitem{b3} F. Wilczek and A. Zee, Phys. Rev. Lett.  {\bf 52}, 2111 (1984).

\bibitem{zanardi} P. Zanardi and M. Rasetti, 
Phys. Lett. A {\bf 264}, 94 (1999).

\bibitem{duan} L. M. Duan, J. I. Cirac, and P. Zoller,
Science {\bf 292}, 1695 (2001).

\bibitem{adiabatic} J. Pachos, P. Zanardi, and M. Rasetti,
Phys. Rev. A {\bf 61}, 010305 (1999).

\bibitem{adiabatic1} A. Recati, T. Calarco, P. Zanardi, J. I. Cirac, and P. Zoller,
Phys. Rev. A {\bf 66}, 032309 (2002).

\bibitem{adiabatic2}    L. Faoro, J. Siewert, and R. Fazio, Phys. Rev. Lett. {\bf 90}, 028301 (2003).

\bibitem{adiabatic3} L.-A. Wu, P. Zanardi, and D. A. Lidar, Phys. Rev. Lett. {\bf 95}, 130501 (2005).

\bibitem{adiabatic4}    P. Zhang, Z. D. Wang, J. D. Sun, and C. P. Sun, Phys. Rev. A {\bf 71}, 042301 (2005).

\bibitem{xdzhang} X. D. Zhang, Q. Zhang, and Z. D. Wang, Phys. Rev. A {\bf 74}, 034302 (2006).

\bibitem{adiabatic5}    I. Kamleitner, P. Solinas, C. M\"{u}ller, A. Shnirman, and M. M\"{o}tt\"{o}nen, Phys. Rev. B {\bf 83}, 214518 (2011).

\bibitem{adiabatic6} V. V. Albert, C. Shu, S. Krastanov, C. Shen, R.-B. Liu, Z.-B. Yang, R. J. Schoelkopf, M. Mirrahimi, M. H. Devoret, and L. Jiang, Phys. Rev. Lett. {\bf 116}, 140502 (2016).


\bibitem{xbwang} W. Xiang-Bin and M. Keiji, Phys. Rev. Lett. {\bf 87}, 097901 (2001).

\bibitem{zhu}    S.-L. Zhu and Z. D. Wang, Phys. Rev. Lett. {\bf 89}, 097902 (2002).

\bibitem{b2} Y. Aharonov and J. Anandan, Phys. Rev. Lett. {\bf 58}, 1593 (1987).

\bibitem{Sjoqvist2012}
E.~Sj\"{o}qvist, D.~M. Tong, L.~{Mauritz Andersson}, B.~Hessmo, M.~Johansson,
  and K.~Singh,
New J.   Phys. \textbf{14}, 103035 (2012).

\bibitem{Xu2012}
G.~F. Xu, J.~Zhang, D.~M. Tong, E.~Sj\"{o}qvist, and L.~C. Kwek,
Phys. Rev. Lett. \textbf{109}, 170501 (2012).


\bibitem{vam2014} V. A. Mousolou and E. Sj\"{o}qvist,
Phys. Rev. A {\bf 89}, 022117 (2014).

\bibitem{Zhang2014d}
J.~Zhang, L.-C. Kwek, E.~Sj\"{o}qvist, D.~M. Tong, and P.~Zanardi,
Phys. Rev. A \textbf{89}, 042302 (2014).

\bibitem{Xu2014}
G.-F. Xu and G.-L. Long,
Sci. Rep. \textbf{4}, 6814 (2014).

\bibitem{Xu2015}
G.~F. Xu, C.~L. Liu, P.~Z. Zhao, and D.~M. Tong,
Phys. Rev. A   \textbf{92}, 052302 (2015).


\bibitem{Xue2015b}
Z.-Y. Xue, J.~Zhou, and Z.~D. Wang,
Phys. Rev. A   \textbf{92}, 022320 (2015).

\bibitem{Xue2016}
Z.-Y. Xue, J.~Zhou, Y.-M. Chu, and Y.~Hu,
Phys. Rev. A \textbf{94},   022331 (2016).

\bibitem{Zhao2016}
P.~Z. Zhao, G.~F. Xu, and D.~M. Tong,
Phys. Rev. A \textbf{94}, 062327 (2016).

\bibitem{Herterich2016}
E.~Herterich and E.~Sj\"{o}qvist,
Phys. Rev. A \textbf{94}, 052310   (2016).


\bibitem{vam2017} V. A. Mousolou,
EPL {\bf 117}, 10006 (2017).

\bibitem{xu2017} G. F. Xu, P. Z. Zhao, T. H. Xing, E. Sj\"{o}qvist, and D. M. Tong, 
    Phys. Rev. A {\bf 95}, 032311 (2017).

\bibitem{Xue2017}
Z.-Y. Xue, F.-L. Gu, Z.-P. Hong, Z.-H. Yang, D.-W. Zhang, Y. Hu, and J. Q. You, 
Phys. Rev. Appl. {\bf 7}, 054022 (2017).

\bibitem{xu20172}
G. F. Xu, P. Z. Zhao, D. M. Tong, and E. Sj\"{o}qvist,
Phys. Rev. A {\bf 95}, 052349 (2017).

\bibitem{zhao2017}
P. Z. Zhao, G. F. Xu, Q. M. Ding, E. Sj\"{o}qvist, and D. M. Tong,
Phys. Rev. A {\bf 95}, 062310 (2017).

\bibitem{vam2016} V. A. Mousolou,
Phys. Rev. A {\bf 96}, 012307 (2017).


\bibitem{Abdumalikov2013}
A.~A. Abdumalikov, J.~M. Fink, K.~Juliusson, M.~Pechal, S.~Berger, A.~Wallraff,
  and S.~Filipp,
Nature (London) \textbf{496}, 482 (2013). 

\bibitem{Feng2013}
G.~Feng, G.~Xu, and G.~Long,
Phys. Rev. Lett. \textbf{110}, 190501   (2013).


\bibitem{li2017}  H. Li, L. Yang, and G. Long,
Sci. China: Phys., Mech. Astron. {\bf 60}, 080311(2017).


\bibitem{Zu2014}
C.~Zu, W.-B. Wang, L.~He, W.-G. Zhang, C.-Y. Dai, F.~Wang, and L.-M. Duan,
Nature (London) \textbf{514}, 72 (2014).   

\bibitem{Arroyo-Camejo2014}
S.~Arroyo-Camejo, A.~Lazariev, S.~W. Hell, and G.~Balasubramanian,
Nat. Commun. \textbf{5}, 4870 (2014).


\bibitem{nv2017} Y. Sekiguchi, N. Niikura, R. Kuroiwa, H. Kano, and H. Kosaka,
Nat. Photonics {\bf 11}, 309 (2017). 

\bibitem{nv20172}  B. B. Zhou, P. C. Jerger, V. O. Shkolnikov, F. J. Heremans, G. Burkard, and D. D. Awschalom, 
Phys. Rev. Lett.  {\bf 119}, 140503 (2017).

\bibitem{sq1} Y. Makhlin, G. Sch\"{o}n, and A. Shnirman, Rev. Mod. Phys. {\bf 73},
357 (2001).

\bibitem{sq2} J. Clarke and F. K.Wilhelm, Nature (London) {\bf 453}, 1031 (2008).

\bibitem{sq3} J. Q. You and F. Nori, Nature (London) {\bf 474}, 589 (2011).

\bibitem{sq4} M. H. Devoret and R. J. Schoelkopf, Science {\bf 339}, 1169- (2013).

\bibitem{transmon} J.  Koch,  T. M. Yu, J. Gambetta, A. A. Houck, D. I. Schuster, J. Majer, A. Blais, M. H. Devoret, S. M. Girvin, and R. J. Schoelkopf,
Phys. Rev. A {\bf 76}, 042319 (2007).

\bibitem{multilevel} M. J. Peterer, S. J. Bader, X. Jin, F. Yan, A. Kamal, T. J. Gudmundsen, P. J. Leek, T. P. Orlando, W. D. Oliver, and S. Gustavsson,
Phys. Rev. Lett. {\bf 114}, 010501 (2015).


\bibitem{jc} S. Zeytino\u{g}lu, M. Pechal, S. Berger, A. A. Abdumalikov Jr., A. Wallraff, and S. Filipp, 
    Phys. Rev. A {\bf 91}, 043846 (2015).



\bibitem{flux1} J. H. Plantenberg, P. C. de Groot, C. J. Harmans, and J. E.
Mooij, Nature (London) {\bf 447}, 836 (2007).

\bibitem{flux2} S. H. W. van der Ploeg, A. Izmalkov, A. M. van den Brink,
U. H\"{u}bner, M. Grajcar, E. Il¡¯ichev, H. G. Meyer, and A. M.
Zagoskin, Phys. Rev. Lett. {\bf 98}, 057004 (2007).

%
\bibitem{lattice} L. Steffen, Y. Salathe, M. Oppliger, P. Kurpiers, M. Baur, C. Lang, C. Eichler, G. Puebla-Hellmann, A. Fedorov, and A. Wallraff,
Nature (London) {\bf 500}, 319 (2013).  

\bibitem{barends} R. Barends, \emph{et al}. 
Nature (London) {\bf 508}, 500 (2014).

\bibitem{yale1} T. Brecht, W. Pfaff, C. Wang, Y. Chu, L. Frunzio, M. H. Devoret,
and R. J. Schoelkopf, npj Quantum Inf. {\bf 2}, 16002 (2016).

\bibitem{yale2} Z. K. Minev, K. Serniak, I. M. Pop, Z. Leghtas, K. Sliwa, M. Hatridge, L. Frunzio, R. J. Schoelkopf, and M. H. Devoret,
Phys. Rev. Appl. {\bf 5}, 044021 (2016).


\bibitem{noise} Y.-P. Wang, W.-L. Yang, Y. Hu, Z.-Y. Xue, and Y. Wu, npj
Quantum Inf. {\bf 2}, 16015 (2016).

\end{thebibliography}
\end{document}